\documentclass[twocolumn]{aastex63}
\usepackage{verbatim}
\usepackage[utf8]{inputenc}
\usepackage{cellspace, longtable}
\usepackage{ulem}
\usepackage{amssymb}
\usepackage{amsmath}

\usepackage{array}

\newcommand{\Ha}{H$\rm \alpha$}
\newcommand{\msun}{$\rm M_\odot~$}

\newcommand{\kms}{\rm km\ s^{-1}}
\newcommand{\ergs}{\rm erg\ s^{-1}}
\newcommand{\oiii}{{\sc [O iii]}}
\newcommand{\oii}{{\sc [O ii]}}

\newcommand{\lya}{Ly$\alpha$} 

\accepted{September 29, 2021}

\shorttitle{Radiation-Driven Emission-line Wings}
\shortauthors{Komarova et al.}

\begin{document}

\title{Emission-line Wings Driven by Lyman Continuum in the Green Pea Analog Mrk 71}

\correspondingauthor{Lena Komarova}
\email{komarova@umich.edu}

\author[0000-0002-5235-7971]{Lena Komarova}
\affiliation{Astronomy Department, University of Michigan,
Ann Arbor, MI, 48103, USA }

\author[0000-0002-5808-1320]{M. S. Oey}
\affiliation{Astronomy Department, University of Michigan,
Ann Arbor, MI, 48103, USA }

\author[0000-0003-3893-854X]{Mark R. Krumholz}
\affiliation{Research School of Astronomy and Astrophysics, Australian National University, Canberra, ACT 2611, Australia}
\author[0000-0002-3814-5294]{Sergiy Silich}
\affiliation{Instituto Nacional de Astrofísica Óptica y Electrónica, AP 51, 72000 Puebla, México}
\author[0000-0002-5320-2568]{Nimisha Kumari}
\affiliation{AURA for ESA, Space Telescope Science Institute, 3700 San Martin Drive, MD 21218, USA}
\author[0000-0003-4372-2006]{Bethan L. James}
\affiliation{AURA for ESA, Space Telescope Science Institute, 3700 San Martin Drive, MD 21218, USA}

\begin{abstract}
We propose that the origin of faint, broad emission-line wings in the Green Pea (GP) analog Mrk~71 is a clumpy, LyC and/or Ly$\alpha$-driven superwind. Our spatially-resolved analysis of Gemini-N/GMOS-IFU observations shows that these line wings with terminal velocity $>3000\ \kms$ originate from the super star cluster (SSC) Knot~A, and propagate to large radii. The object's observed ionization parameter and stellar surface density are close to their theoretical maxima, and radiation pressure dominates over gas pressure. Together with a lack of evidence for supernova feedback, these imply a radiation-dominated environment. We demonstrate that a clumpy, radiation-driven
superwind from Knot A is a viable model for generating the extreme velocities, and in particular, that Lyman continuum and/or Ly$\alpha$ opacity must be responsible.
We find that the Mrk 71 broad wings are best fitted with power laws, as are those of a representative extreme GP and a luminous blue variable star, albeit with different slopes.
This suggests that they may share a common wind-acceleration mechanism. We propose that high-velocity, power-law wings may be a distinctive signature of radiation feedback, and of radiatively-driven winds, in particular.
\end{abstract}


\section{Introduction}
\label{sec:intro}

Mrk~71 is a starburst in the nearby, metal-poor \citep[$12+\log\rm O/H=7.89;$][]{Izotov1997} galaxy NGC 2366.  
This system is the nearest \citep[3.4 Mpc;][]{Tolstoy1995} analog of the Green Peas (GPs), a class of local ($z\approx$ 0.2) high-ionization parameter galaxies of intense cosmological interest due to high incidence of Lyman-continuum (LyC) escape \cite[e.g.,][]{Izotov2016,Izotov2018}.
The properties of Mrk~71/NGC~2366 are fully consistent with those of GPs in terms of ionization parameter, specific star formation rate, metallicity, and parameters linked to low LyC optical depth \citep{Micheva2017}.  

\begin{figure*}
\gridline{\fig{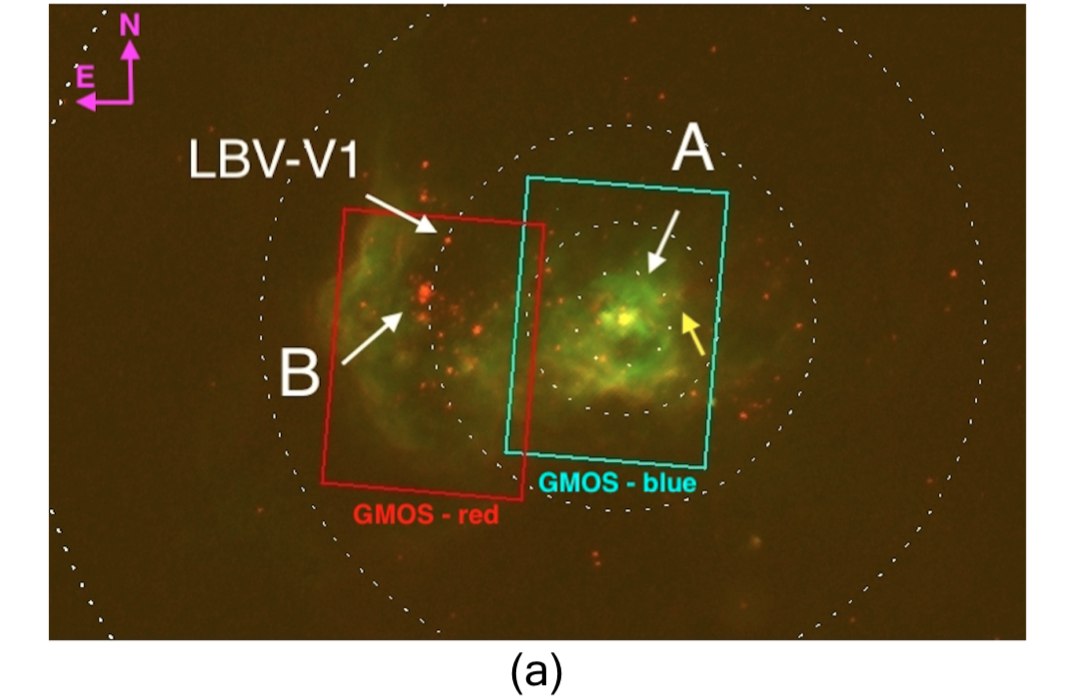}{0.5\textwidth}{}
            \fig{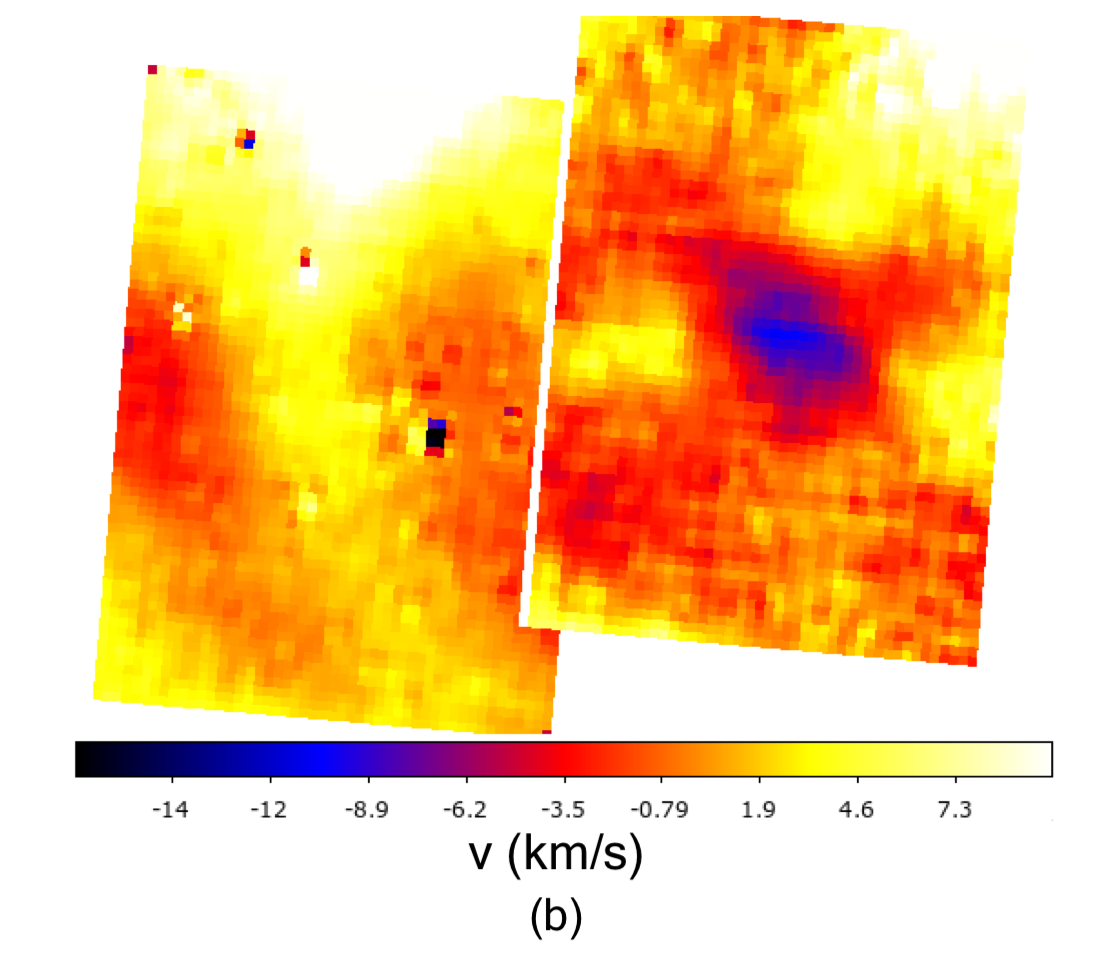}{0.5\textwidth}{}
          }
\caption{(a) HST/WFC3 F373N (\oii, red) and continuum-subtracted F502N (\oiii, green) two-color image
of Mrk 71, showing the GMOS-IFU footprints. Knots A and B are indicated, and the yellow arrow highlights elephant trunks.  The dotted lines show radii of 20 (CB), 40, 80, 150, and 250 pc. LBV-V1 is a luminous blue variable in the system. (b) Velocity map in \Ha~(east) and \oiii$\lambda$4959 (west) from GMOS-IFU data.}
\label{fig:Haimages}
\end{figure*}

Mrk~71 is dominated by Knot~A, a super star cluster (SSC) that is still enshrouded by its natal gas (Figure~\ref{fig:Haimages}a), yet dominates the galaxy’s nebular luminosity and ionization parameter. 
This suggests that mechanical feedback from stellar winds and supernovae is ineffective, as expected in 
metal-poor systems, where stellar winds are weak (e.g., \citealt{vink2001}, \citealt{Ramachandran}) 
and the onset of supernovae delayed (e.g., \citealt{Heger2003}, \citealt{Sukhbold}).  Compounded by high gas densities and pressures, these conditions lead to catastrophic cooling that suppresses adiabatic, cluster-driven winds  \cite[e.g.,][]{Silich2004,Silich2017, Lochhaas2017}. 
This scenario apparently applies in several M82 SSCs \citep{Smith2006, Westmoquette}, NGC 5253 \citep[e.g.,][]{Silich2020}, and a few extreme GPs \citep{Jaskot2017}.
\cite{Oey2017} detect $10^5$ \msun of molecular gas within 7 pc of Knot~A that shows momentum-conserving expansion, implying that
the SSC has failed to clear its environment, unlike its older neighbor, Knot~B (Figure~\ref{fig:Haimages}a). 

A distinctive feature shared by both Mrk71 and extreme GPs is the presence of extremely broad wings in strong nebular emission lines, e.g., \oiii\ and \Ha~\citep{Amorin2012, Izotov2007}. Despite multiple investigations \citep{roy1992,Binette2009}, the origin of these broad (full-width-zero-intensity FWZI $\gtrsim$ 6000 km/s) wings in Mrk~71 remains unknown.
In this Letter, we argue that they are a signature of a clumpy, radiation-driven superwind.

\section{Analysis of Mrk 71 Broad Wings}\label{sec:obs}

We obtained archival  Gemini-N/GMOS-IFU spectral cubes of Mrk 71 Knots A and B. The two-slit mode of the GMOS IFU covers a $5\arcsec \times 7\arcsec$ field of view (FOV), which is sampled by 1000 on-target, 0.2$\arcsec$ lenslets and 500 lenslets for background. Knot A was observed on 2006 December 28, using grating B600 (GN-2006B-Q-65) with an exposure of 1200 s, covering 4090–5400\AA, including \oiii$\lambda \lambda$4959, 5007 at 2.5\AA\ (150 $\kms$) resolution. 
Knot~B was observed on 2008 January 27, using grating R831 (GN-2007B-Q-44) with exposure $2\times 1200$~s, 
covering 6025–6760\AA, including \Ha\ at 1.9\AA\ (87 $\kms$) resolution. We reduced both sets of observations using the standard GEMINI pipeline in  \textsc{iraf},\footnote{IRAF was distributed by NOAO, which was managed by AURA under a cooperative agreement with NSF. }  including bias subtraction, flat-field correction, wavelength calibration and sky subtraction.  
Our final data cubes have spatial sampling of 0.1$\arcsec$ spaxel$^{-1}$, and are normalized to the systemic velocity of 79 $\kms$ for this region of NGC 2366 \citep{Hunter2001}. For more details on data reduction, see \citet{Kumari2018}. 
\oiii$\lambda$5007 falls on a chip gap, and so we use \oiii$\lambda4959$ for all measurements.

First examining the line core kinematics, Figure~\ref{fig:Haimages}b shows blueshifted gas around Knot~A coinciding with  a more optically thin region, the ``Crystal Ball'' (CB; innermost dotted line in Figure~\ref{fig:Haimages}a), surrounded by denser gas. We adopt $2.8\times10^5$ \msun for the SSC mass (\citealt{Gonzalez-delgado1994}, hereafter GD94; \citealt{Micheva2017}), implying an expected stellar wind mechanical luminosity 
$L_{\rm w}\sim 10^{39}\ \ergs$ \citep{Oey2017} for the Mrk71 metallicity. 
Adopting a SSC age of 1 Myr and a mean density of 400 cm$^{-3}$, a momentum-conserving wind bubble model predicts a radius of 20 pc and a velocity of $\sim 9.8\ \kms$ \citep[e.g.,][]{McCraySnow1979}, consistent with the observed CB properties (Figure~\ref{fig:Haimages}b).

\begin{figure*}
\centering
\gridline{\fig{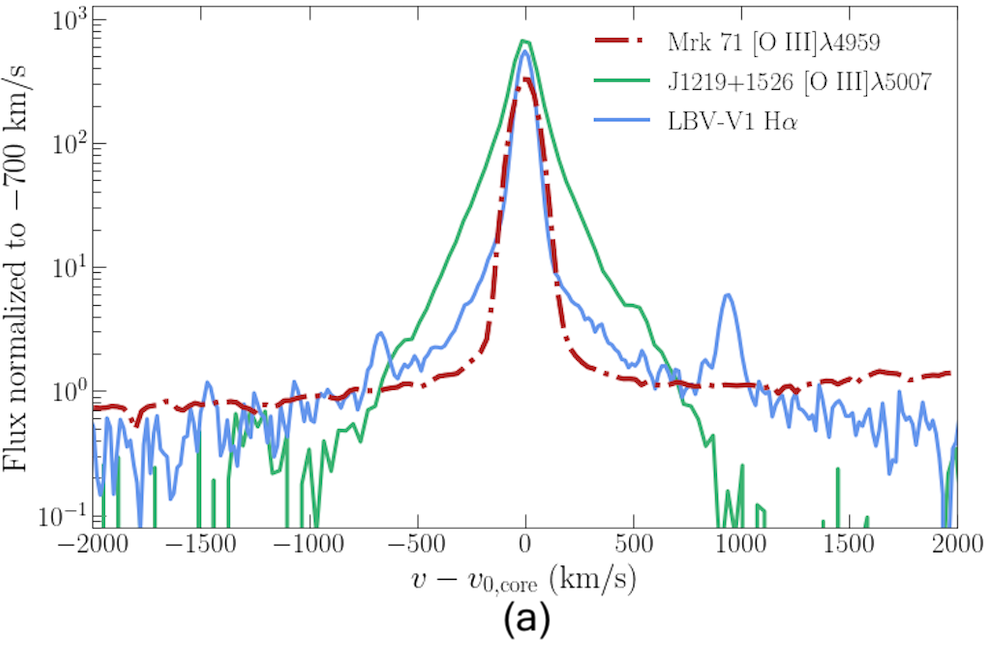}{0.5\textwidth}{}
            \fig{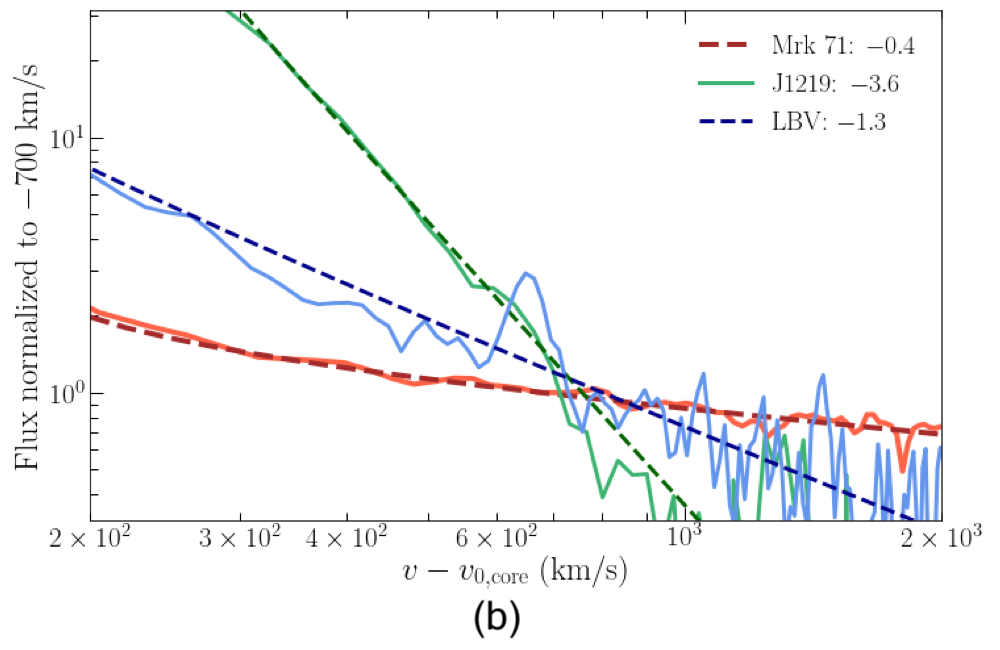}{0.5\textwidth}{}}
\caption{(a) Comparison of emission-line profiles for Mrk 71, LBV-V1, and GP J1219+1526, recentered at the best-fit line core centroids $v_{0,\rm core}$. The profile of J1219+1526 is Doppler-corrected for z = 0.196. (b) Power law models fitted to the blue broad wings for each object in panel (a). Fitted slopes are shown in the legend. }
\label{fig:GPs}
\end{figure*}

The broad-wing component has $v_\infty > 3000\ \kms$, and 
to isolate it from the emission-line cores, we carry out non-linear least squares fits to the continuum-subtracted line profiles of \Ha, $\rm H\beta$, and \oiii$\lambda$4959. We use a functional form consisting of a Gaussian core and a power-law wing of the form $A(v-v_{\rm0,wing})^p$, where $v_{\rm 0,wing}$ is the velocity centroid of the broad wing component. We show our best fits in Figure~\ref{fig:GPs}. The Mrk 71 \oiii$\lambda$4959 profile shown is spatially integrated over the west GMOS FOV. The core contributes $< 2$\% at $|v|>240\ \kms$, so we take $|v| > 200\ \kms$ to mark the core-wing transition. We note that it cannot be well-fit by
multiple gaussians.

\begin{figure*}
\gridline{\fig{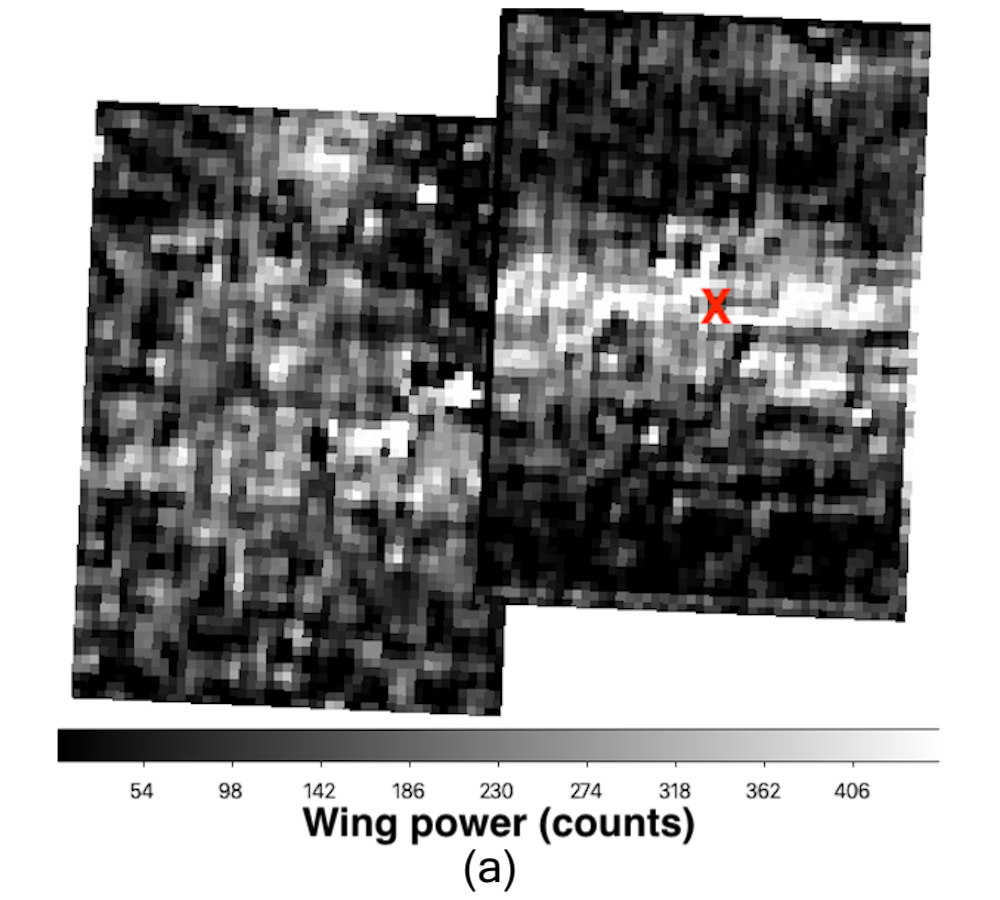}{0.33\textwidth}{}
           \fig{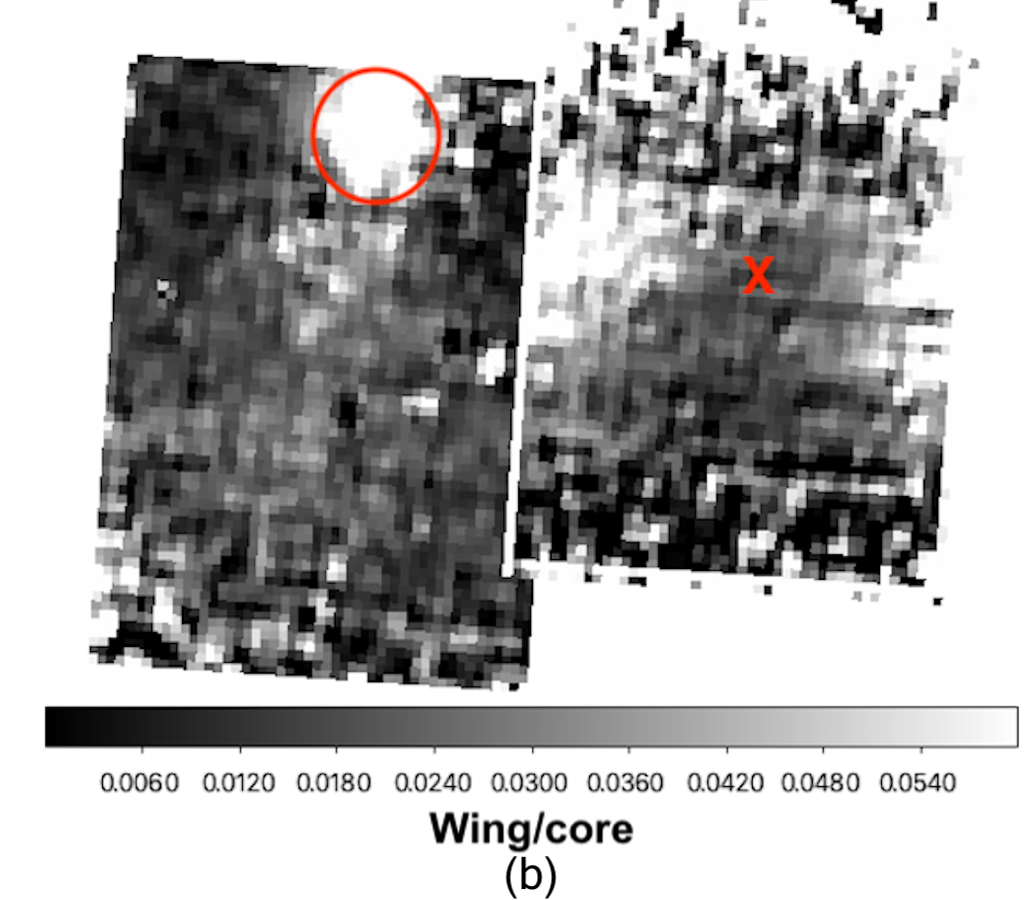}{0.33\textwidth} {}
          \fig{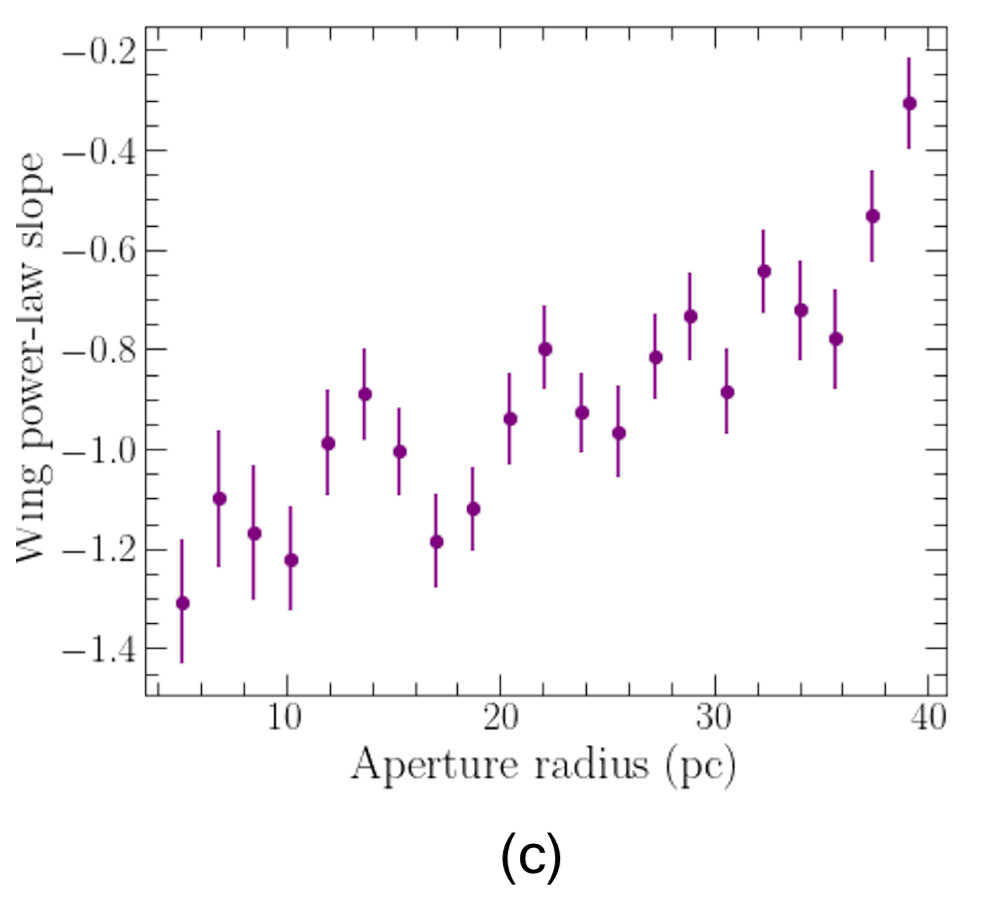}{0.33\textwidth}{}}
\caption{Spatially resolved broad wing properties calculated spaxel-by-spaxel in \Ha\ (east) and \oiii$\lambda4959$ (west) from GMOS data cubes. Knot A is marked with a red cross.
(a) Broad wing emission for $|v|> 200\ \kms$ (see text); (b) Ratio of total broad wing emission to line core ($\vert v \vert$  \textless 200 km/s). The aperture used for LBV-V1 is shown as a red circle.
(c) Power-law slopes fitted to the broad wings in annuli centered on Knot A, as a function of aperture radius in west cube. 
\label{fig:spatialwings}}
\end{figure*}

Figure \ref{fig:spatialwings}a maps the wing flux, summed in velocity,
where \oiii$\lambda$4959 or \Ha~emission is detected 
at least 1$\sigma$ above continuum. The wing flux is elevated around Knot A, but stays mostly uniform out to large radii, while the narrow component emission decreases with distance from Knot~A. \citet{roy1991} and GD94 show that the broad wing emission extends over a region $\gtrsim 250$ pc in radius, well beyond the GMOS FOV.  Based on its surface brightness profile (GD94), we adopt a characteristic broad wing length scale $R_{\rm broad} = 250$ pc.

Figure~\ref{fig:spatialwings}b maps the ratio of broad wing to line core emission.
The observed ratio is 1--5\% and lowest around Knot A, where the line core dominates. The luminous blue variable LBV-V1 \citep[][Figure~\ref{fig:Haimages}a]{DrissenLBV} is also prominent in \Ha~in the northern part of Fig.~\ref{fig:spatialwings}b, since LBVs also have broad Balmer emission-line wings. We take advantage of this in Section \ref{GPlink} to compare the radiatively-driven LBV line profile to that of Mrk 71. Figure~\ref{fig:spatialwings}b suggests that the broad wings and line cores are independent spectral components, and likely kinematically energized through different physical processes. 
The spatial variation reflects the distribution of the dominant, core component.

\begin{figure*}
\includegraphics[width=\textwidth]{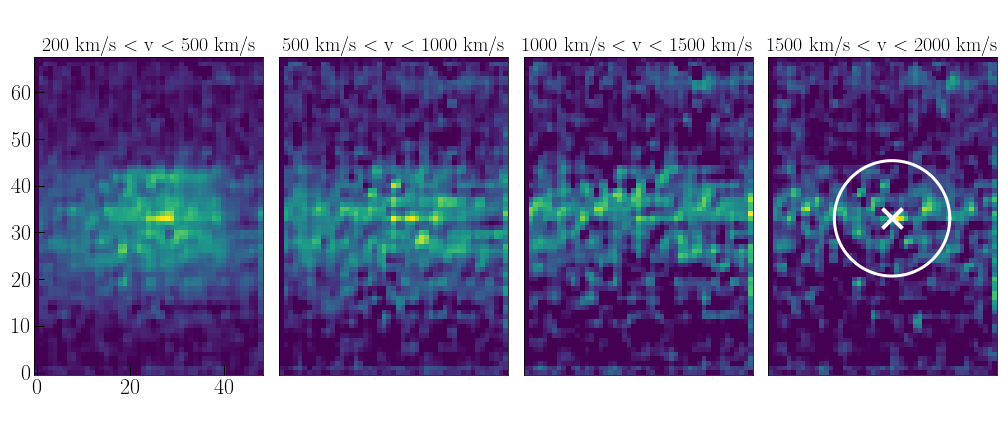}
\caption{Broad wing emission in \oiii$\lambda4959$ from different velocity bins. Axes are in pixels. The color scale shows integrated wing flux in each velocity bin in GMOS counts. Knot A and 20-pc Crystal Ball are marked in the last panel as a cross and a circle, respectively.}
\label{fig:velbins}
\end{figure*}

We also fit the wing emission in annular regions as a function of distance from Knot~A.
Figure~\ref{fig:spatialwings}c shows 
that the broad-wing power-law slope steepens towards Knot A, 
from $p \sim-0.2$ to $-1.5$ from outermost to innermost regions, respectively.
This implies an increasing contribution of low-velocity gas in the innermost regions. 
This is also seen in Figure \ref{fig:velbins}, which maps wing emission by velocity bins. The lowest-velocity material is more prevalent at smaller scales, where the CB structure is apparent,
while the higher-velocity gas has more uniform distribution. 
Thus, Figures~\ref{fig:spatialwings}c and \ref{fig:velbins}
show that {\it the broad wing emission originates from Knot A.}

Using the \Ha~HST image calibrated by \citet{James2015} and a constant  wing/core ratio of 2\%, we 
obtain median broad wing \Ha\ emission measures (EM) 
in annular zones bounded by radii of 40, 80, 150, and 250~pc of 1200, 360, 90, and 21 pc~cm$^{-6}$, respectively.
Using a conversion of $2\times10^{-18} \ergs cm^{-2} arcsec^{-2} = 1~pc~cm^{-6}$, these give broad wing densities of 
1.6, 0.9, 0.4, and 0.3 $\rm cm^{-3}$, respectively, assuming an emitting sphere of radius 250 pc, yielding a total broad wing mass of $M_{\rm bw}=5.5\times10^5$ \msun within a 250-pc-radius sphere. 
However, this is only an upper limit since any unresolved, dense clumps will dominate the EM.

\section{Radiation-driven Feedback} \label{superwindmodel}

First identified by \cite{roy1991}, the faint broad wings in Mrk71 have been a puzzle for decades. 
\citet{Chu1994} find that the broad \Ha\ line profile in 30 Doradus consists of many individual kinematic components whose sum is, per the central limit theorem, a broad Gaussian. However, as discussed above, the line wings in Mrk~71 are not Gaussian. \cite{roy1992} find that electron scattering cannot adequately explain the wing profiles, either.
\cite{Binette2009} propose a model of turbulent mixing layers (TMLs) for Mrk~71 that reproduces the wing profiles reasonably well, but
requires a hot supersonic wind to drive them. Other mechanisms also depend on energy-driven feedback \citep[e.g.,][]{Tenorio-Tagle1997}. However, adiabatic feedback from stellar winds and supernovae do not seem to explain the Mrk71 wings \citep[GD94,][]{roy1992, Oey2017}.

Indeed, several additional lines of evidence argue against adiabatic feedback as an explanation for the wings. The SSC has not been cleared sufficiently to even expose its stellar population \citep{Drissen2000}, as expected for an 
energy-dominated supernova- or wind-driven bubble. XMM-Newton observations show only faint X-rays from Mrk~71 ($L_{\rm X}=8\times 10^{36}\ \ergs$; \citealt{Thuan2014}), and these are spatially associated with Knot~B rather than A. Theoretically, at the metallicity of Mrk~71 we expect stellar winds to be weaker, and the onset of supernovae to be delayed, compared to those in a more metal-rich population. Moreover, the high density and compactness of Knot A imply that, even if stellar winds are launched, they are likely to cool catastrophically, losing a large fraction of the deposited mechanical energy \cite[e.g.,][]{Silich2004,Tenorio-Tagle2007, Lochhaas2017}.

Instead, we propose that the fast winds are radiatively driven \citep{Ishibashi, Thompson}. Several lines of evidence favor the importance of radiation pressure in this system. Given the stellar mass $M_* = 2.8\times 10^5$ M$_\odot$ and the observed radius of Knot A, $r_{\rm c} \approx 1.5$ pc, the stellar surface density is close to the upper limit of $\approx 10^5$ M$_\odot$ pc$^{-2}$ at which radiation pressure on dust grains is expected to disrupt the system \citep{Crocker2018}. The mean radiation pressure inside radius $r$ is $P_{\rm rad} = 3 L_*/4\pi r^2 c$ \citep{Lopez11a}, where we adopt $L_* \sim 3 \times 10^{41}$ erg s$^{-1}$ for Knot A, since the unresolved 
total IR luminosity of NGC 2366 is $5\times 10^{41}$ erg s$^{-1}$ \citep{Dale2009}. Comparing this to the thermal pressure $P_{\rm th}=2 n k T$ of photoionized gas\footnote{Assuming 2 free particles per free electron.} with electron density $n\sim 400$ cm$^{-3}$ and temperature $T\sim 16,000$ K shows that $P_{\rm rad} > P_{\rm th}$ for all $R \lesssim 12$ pc, as expected for SSCs \citep{Krumholz2009}. 
We also note that \citet{Dopita2002} and \citet{Yeh2012} show that an ionization parameter of $\log U\sim-2$ is a maximum value expected for a system when radiation pressure exceeds ionized gas pressure. For Knot~A, $\log U\approx -1.9$ \citep{Micheva2017}, implying that radiation is dynamically important.

Can radiation pressure explain the broad wings? For radiation pressure on dust grains, as proposed by \citet{Ishibashi} and \citet{Thompson}, the answer is no, for the following reason: the terminal velocity of a radiatively driven wind encountering an optically thick medium is \citep{Krumholz2017}
\begin{equation}
    v_{\infty} = \sqrt{\Gamma \tau_0-1}~v_{\rm esc},
\end{equation}
where $v_{\rm esc}$ is the escape speed from the wind launch region.
Additionally,
\begin{equation}
    \Gamma = \frac{L_*}{4\pi G M_* \Sigma c},
\end{equation}
where
$\Sigma$ is the gas surface density, $\tau_0 = \kappa \Sigma$ is the optical depth of the launching region, and $\kappa$ is the specific opacity. Thus, for Mrk~71 we have
\begin{equation}
    v_{\infty} = \sqrt{\frac{L_* \kappa}{4\pi G M_* c} - 1} \times v_{\rm esc} ~~ \approx 5~\kappa_3^{1/2}~v_{\rm esc},
\end{equation}
where $\kappa_3 = \kappa / 1000$ cm$^2$ g$^{-1}$, and the second approximate equality holds for $\kappa_3 \gtrsim 1$. For the Knot A SSC, $v_{\rm esc} \sim 40$ km s$^{-1}$, so producing the observed $\approx 3000$ km s$^{-1}$ outflow speed requires $\kappa_3 \approx 300$; this is two orders of magnitude higher than the UV dust opacity even at solar metallicity \citep{Draine2003}, and Mrk~71 is a sub-Solar system. Thus dust opacity cannot explain the observed velocities.

However, Mrk~71, like the GPs to which it is analogous, is a candidate LyC emitter. If LyC is able to escape from the vicinity of Knot A to reach large radii, and the material producing the broad wings contains a substantial neutral component, then the relevant opacity is the LyC opacity of neutral hydrogen, which is $\kappa = 2.7\times 10^6$ cm$^2$ g$^{-1}$ at threshold.  This is high enough that, even reducing $\Gamma$ by a factor of $\approx 3$ to account for the fact that only $\approx 1/3$ of the bolometric luminosity of a young stellar population is emitted at $>13.6$ eV, material with a neutral fraction $\geq 30\%$ would reach terminal velocities $v_\infty \geq 3000$ km s$^{-1}$. Moreover, if LyC radiation is escaping to large radii, then \lya\ likely does so as well. Hence, the effective opacity is even higher, though by how much depends sensitively on the wind velocity profile.

Thus {\it acceleration by LyC or \lya\ photons can explain the large observed terminal velocities.} However, the wind must be very clumpy to remain within the allowed momentum budget: assuming an ionizing luminosity $L_{*,\rm ion} = 10^{41}$ erg s$^{-1}$, this provides a total momentum $L_{*,\rm ion} t_*/c \approx 10^6$ M$_\odot$ km s$^{-1}$, where $t_*\sim 1$ Myr is the SSC age; given the typical $\approx 3000$ km s$^{-1}$ speed, we obtain $M_{\rm bw} \approx\rm a\ few\times 10^2$ \msun.  Comparing to our EM-based estimates (Section~\ref{sec:obs}), this in turn requires that the emitting material have mean densities $\sim 10^2-10^3$ cm$^{-3}$, compared to the minimum of $\sim 0.3$ cm$^{-3}$ we estimated assuming a 100\% volume filling fraction.
This explains how the wind, and also UV radiation, can pass through the swept-up CB shell (Figure~\ref{fig:Haimages}b), which must itself be clumpy.

\section{Discussion}

A radiation-driven wind model
is consistent with a catastrophic cooling scenario where adiabatic feedback is suppressed 
within the SSC. Other signatures also indicate radiation-dominated feedback, such as ``elephant trunks'' seen around Knot A (Figure~\ref{fig:Haimages}a) and
the extreme nebular excitation \citep[e.g.,][]{Micheva2017}.
The CB size and morphology is consistent with it having originated as a dense, highly optically thick region, suitable for 
providing the dense knots that accelerate the wind.
This radiation feedback, together with the momentum-conserving, stellar wind-driven, expanding shell (Figure~\ref{fig:Haimages}b), has partially cleared the CB gas.

The resolved kinematic structure in Figure~\ref{fig:velbins} is consistent with radial acceleration. This causes the power-law slope to flatten at larger radii (Figure~\ref{fig:spatialwings}c) where there is less contribution from lower-velocity material. The shape and symmetry of the broad wing profiles can be explained by TMLs driven by this wind \citep{Binette2009}. Projection effects also may be relevant \citep[e.g.,][]{Krumholz2017}.
Thus, the spatial kinematics are qualitatively consistent with a radiation-driven wind.  
However, we cannot rule out some contribution to the early acceleration from stellar winds.

\subsection{Possible Link to Green Peas} \label{GPlink}

Mrk71/NGC 2366 is a critical nearby analog of the Green Peas. Some of the most extreme GPs also show broad, symmetric, emission-line wings, although with lower maximum velocities \citep{Amorin2012, Hogarth2020}.
Our result that this feature is driven by LyC/Ly$\alpha$ suggests a possible similar process in these objects, consistent with their relatively high LyC/Ly$\alpha$ transparency.

As part of a larger study of GP line profiles, we obtained a MIKE/Magellan echelle spectrum of the GP J1219+1526, which exhibits among the most extreme nebular excitations \citep{Jaskot2014, Ravindranath}. The spectra were taken 2015 January 12 and reduced with the CarPy pipeline \citep{Kelson2000}. This object is one of the GPs exhibiting mostly symmetric non-Gaussian profiles and broad wings.
We tried
multiple-Gaussian, Lorentzian, power-law, and exponential distributions and find that, while
the core of the \oiii$\lambda5007$ profile of J1219+1526 is best fit by a Lorentzian, its broad wings are power laws with slope $\sim -4$.
We also compare with the \Ha\ line profile for LBV-V1, since LBV winds are also radiation-driven \citep[e.g.,][]{Puls2008review}.  Figure~\ref{fig:spatialwings}b shows the extraction aperture for this star.
Figure \ref{fig:GPs} shows that Mrk 71, J1219+1526, and LBV-V1 all have power-law wings, suggesting that the underlying physical mechanism may be similar in all three objects.
Active galactic nuclei show similar line wings and also may have radiation-driven outflows \citep[e.g.,][]{Thompson}.

The power-law slopes vary between the three objects, which may be due to different opacities, acceleration laws, and viewing angles \citep{Krumholz2017}.
In particular, mass-loading may be important in setting the power-law slope.  Mass-loading is likely important, since the optically thick clumps driving the acceleration are subject to ablation and evaporation.  The radial dependence of this process is a function of the large-scale density, structure, and metallicity of the parent environment.

We argue that smooth, symmetric, power-law wings, as exhibited by Mrk 71 and J1219+1526 are signatures of radiation-driven winds, where nebular emission originates from a filled volume of small clumps, thus generating a smooth profile in both space and velocity.  This is seen in the 2D \Ha\ spectrum of Mrk 71 from \cite{Gonzalez-delgado1994}. Adiabatic feedback, by contrast, is more likely to produce multiple nebular components associated with shells, as seen in, e.g., 30 Doradus \citep{Chu1994}, which shows nebular line splitting and asymmetry. Indeed, many GP line profiles are better fit by multiple Gaussian components \citep{Amorin2012,Hogarth2020, Bosch2019}, which appear to be unresolved, individual star-forming complexes with outflows driven by stellar winds and SNe. 
Thus, some GP systems may be radiation-dominated, while most may conform to the classic adiabatic model. As the wings observed in Mrk 71 are 100$\times$ fainter than the line cores, lower S/N in GP spectra also may 
inhibit detection of the wings in many objects. Further study is needed to confirm the existence and frequency of radiation-driven winds in GPs.

However, radiation-dominated feedback in GPs can be expected, since by definition these are compact, extreme starbursts with exceptional ionization parameters, which \citet{Dopita2002} directly link to radiation-driven flows.  The low  GP optical depths in Ly$\alpha$ and LyC are themselves evidence of these conditions. 

Thus Mrk~71, and perhaps some extreme GPs, are likely dominated by
radiation feedback, promoted by low metallicity and high star-formation rate density.
Broad, power-law, emission-line wings may originate from radiation-driven superwinds, and may indicate those conditions. 
Such a model is compatible with observations of Mrk~71. The high $v_\infty > 3000\ \kms$ appears to be driven by LyC and \lya\ opacity, which is fully consistent with its status as a GP analog.

\section{Conclusions}

Our analysis of GMOS-N IFU data shows that broad (FWZI $> 6000\ \kms$), power-law emission-line wings observed in Mrk 71 originate from Knot A, and the spatially-resolved kinematics are consistent with a wind velocity structure. We argue that the broad wings do not originate from stellar winds or supernovae, but are instead radiation-driven, based on the ionization parameter, threshold stellar surface density, and radiation pressure dominance.

We show that a radiation-driven superwind can explain these extreme velocities. The required opacity can only be provided by LyC and/or Ly$\alpha$ acting on dense, optically thick clumps. We find that an example extreme GP and LBV also show power-law line wings, suggesting that
this feature may serve as a signature of radiation-driven winds and feedback. Since such a wind requires \lya/LyC photons to reach large distances, it also could be a signature of LyC escape.

\acknowledgments
We thank Laurent Drissen and Mike Irwin for discussions and information regarding the GMOS-IFU observations and data reduction.  We are also grateful to Laura Lopez for a useful science discussion. We thank the anonymous referee for useful suggestions. This work was supported by HST-GO-14080 and HST-GO-16261 to M.S.O. M.R.K. acknowledges support from the Australian Research Council through its \textit{Discovery Projects} and \textit{Future Fellowship} funding schemes, awards DP190101258 and FT180100375. S.S. acknowledges support from CONACYT-México through research grant A1-S-28458. 

\bibliography{paper_clean}{}

\begin{thebibliography}{}
\expandafter\ifx\csname natexlab\endcsname\relax\def\natexlab#1{#1}\fi
\providecommand{\url}[1]{\href{#1}{#1}}
\providecommand{\dodoi}[1]{doi:~\href{http://doi.org/#1}{\nolinkurl{#1}}}
\providecommand{\doeprint}[1]{\href{http://ascl.net/#1}{\nolinkurl{http://ascl.net/#1}}}
\providecommand{\doarXiv}[1]{\href{https://arxiv.org/abs/#1}{\nolinkurl{https://arxiv.org/abs/#1}}}

\bibitem[{{Amor{\'\i}n} {et~al.}(2012){Amor{\'\i}n}, {V{\'\i}lchez},
  {H{\"a}gele}, {Firpo}, {P{\'e}rez-Montero}, \& {Papaderos}}]{Amorin2012}
{Amor{\'\i}n}, R., {V{\'\i}lchez}, J.~M., {H{\"a}gele}, G.~F., {et~al.} 2012,
  \apjl, 754, L22, \dodoi{10.1088/2041-8205/754/2/L22}

\bibitem[{Binette {et~al.}(2009)Binette, Drissen, {\'{U}}beda, Raga, Robert, \&
  Krongold}]{Binette2009}
Binette, L., Drissen, L., {\'{U}}beda, L., {et~al.} 2009, \aap, 500, 817,
  \dodoi{10.1051/0004-6361/200811132}

\bibitem[{{Bosch} {et~al.}(2019){Bosch}, {H{\"a}gele}, {Amor{\'\i}n}, {Firpo},
  {Cardaci}, {V{\'\i}lchez}, {P{\'e}rez-Montero}, {Papaderos}, {Dors},
  {Krabbe}, \& {Campuzano-Castro}}]{Bosch2019}
{Bosch}, G., {H{\"a}gele}, G.~F., {Amor{\'\i}n}, R., {et~al.} 2019, \mnras,
  489, 1787, \dodoi{10.1093/mnras/stz2230}

\bibitem[{{Chu} \& {Kennicutt}(1994)}]{Chu1994}
{Chu}, Y.-H., \& {Kennicutt}, Robert~C., J. 1994, \apj, 425, 720,
  \dodoi{10.1086/174017}

\bibitem[{{Crocker} {et~al.}(2018){Crocker}, {Krumholz}, {Thompson},
  {Baumgardt}, \& {Mackey}}]{Crocker2018}
{Crocker}, R.~M., {Krumholz}, M.~R., {Thompson}, T.~A., {Baumgardt}, H., \&
  {Mackey}, D. 2018, \mnras, 481, 4895, \dodoi{10.1093/mnras/sty2659}

\bibitem[{{Dale} {et~al.}(2009){Dale}, {Cohen}, {Johnson}, {Schuster},
  {Calzetti}, {Engelbracht}, {Gil de Paz}, {Kennicutt}, {Lee}, {Begum},
  {Block}, {Dalcanton}, {Funes}, {Gordon}, {Johnson}, {Marble}, {Sakai},
  {Skillman}, {van Zee}, {Walter}, {Weisz}, {Williams}, {Wu}, \&
  {Wu}}]{Dale2009}
{Dale}, D.~A., {Cohen}, S.~A., {Johnson}, L.~C., {et~al.} 2009, \apj, 703, 517,
  \dodoi{10.1088/0004-637X/703/1/517}

\bibitem[{{Dopita} {et~al.}(2002){Dopita}, {Groves}, {Sutherland}, {Binette},
  \& {Cecil}}]{Dopita2002}
{Dopita}, M.~A., {Groves}, B.~A., {Sutherland}, R.~S., {Binette}, L., \&
  {Cecil}, G. 2002, \apj, 572, 753, \dodoi{10.1086/340429}

\bibitem[{{Draine}(2003)}]{Draine2003}
{Draine}, B.~T. 2003, \araa, 41, 241,
  \dodoi{10.1146/annurev.astro.41.011802.094840}

\bibitem[{{Drissen} {et~al.}(1997){Drissen}, {Roy}, \& {Robert}}]{DrissenLBV}
{Drissen}, L., {Roy}, J.-R., \& {Robert}, C. 1997, \apjl, 474, L35,
  \dodoi{10.1086/310417}

\bibitem[{Drissen {et~al.}(2000)Drissen, Roy, Robert, Devost, \&
  Doyon}]{Drissen2000}
Drissen, L., Roy, J.-R., Robert, C., Devost, D., \& Doyon, R. 2000, \aj, 119,
  688, \dodoi{10.1086/301204}

\bibitem[{{Gonzalez-Delgado} {et~al.}(1994){Gonzalez-Delgado}, {Perez},
  {Tenorio-Tagle}, {Vilchez}, {Terlevich}, {Terlevich}, {Telles},
  {Rodriguez-Espinosa}, {Mas-Hesse}, {Garcia-Vargas}, {Diaz}, {Cepa}, \&
  {Castaneda}}]{Gonzalez-delgado1994}
{Gonzalez-Delgado}, R.~M., {Perez}, E., {Tenorio-Tagle}, G., {et~al.} 1994,
  \apj, 437, 239, \dodoi{10.1086/174992}

\bibitem[{{Heger} {et~al.}(2003){Heger}, {Fryer}, {Woosley}, {Langer}, \&
  {Hartmann}}]{Heger2003}
{Heger}, A., {Fryer}, C.~L., {Woosley}, S.~E., {Langer}, N., \& {Hartmann},
  D.~H. 2003, \apj, 591, 288, \dodoi{10.1086/375341}

\bibitem[{{Hogarth} {et~al.}(2020){Hogarth}, {Amor{\'\i}n}, {V{\'\i}lchez},
  {H{\"a}gele}, {Cardaci}, {P{\'e}rez-Montero}, {Firpo}, {Jaskot}, \&
  {Ch{\'a}vez}}]{Hogarth2020}
{Hogarth}, L., {Amor{\'\i}n}, R., {V{\'\i}lchez}, J.~M., {et~al.} 2020, \mnras,
  494, 3541, \dodoi{10.1093/mnras/staa851}

\bibitem[{Hunter {et~al.}(2001)Hunter, Elmegreen, \& van Woerden}]{Hunter2001}
Hunter, D.~A., Elmegreen, B.~G., \& van Woerden, H. 2001, \apj, 556, 773–800,
  \dodoi{10.1086/321611}

\bibitem[{{Ishibashi} \& {Fabian}(2015)}]{Ishibashi}
{Ishibashi}, W., \& {Fabian}, A.~C. 2015, \mnras, 451, 93,
  \dodoi{10.1093/mnras/stv944}

\bibitem[{{Izotov} {et~al.}(2016){Izotov}, {Schaerer}, {Thuan}, {Worseck},
  {Guseva}, {Orlitov{\'a}}, \& {Verhamme}}]{Izotov2016}
{Izotov}, Y.~I., {Schaerer}, D., {Thuan}, T.~X., {et~al.} 2016, \mnras, 461,
  3683, \dodoi{10.1093/mnras/stw1205}

\bibitem[{{Izotov} {et~al.}(2007){Izotov}, {Thuan}, \& {Guseva}}]{Izotov2007}
{Izotov}, Y.~I., {Thuan}, T.~X., \& {Guseva}, N.~G. 2007, \apj, 671, 1297,
  \dodoi{10.1086/522923}

\bibitem[{{Izotov} {et~al.}(1997){Izotov}, {Thuan}, \&
  {Lipovetsky}}]{Izotov1997}
{Izotov}, Y.~I., {Thuan}, T.~X., \& {Lipovetsky}, V.~A. 1997, \apjs, 108, 1,
  \dodoi{10.1086/312956}

\bibitem[{{Izotov} {et~al.}(2018){Izotov}, {Worseck}, {Schaerer}, {Guseva},
  {Thuan}, {Fricke}, \& {Orlitov{\'a}}}]{Izotov2018}
{Izotov}, Y.~I., {Worseck}, G., {Schaerer}, D., {et~al.} 2018, \mnras, 478,
  4851, \dodoi{10.1093/mnras/sty1378}

\bibitem[{James {et~al.}(2015)James, Auger, Aloisi, Calzetti, \&
  Kewley}]{James2015}
James, B.~L., Auger, M., Aloisi, A., Calzetti, D., \& Kewley, L. 2015, \apj,
  816, 40, \dodoi{10.3847/0004-637x/816/1/40}

\bibitem[{{Jaskot} \& {Oey}(2014)}]{Jaskot2014}
{Jaskot}, A.~E., \& {Oey}, M.~S. 2014, \apjl, 791, L19,
  \dodoi{10.1088/2041-8205/791/2/L19}

\bibitem[{{Jaskot} {et~al.}(2017){Jaskot}, {Oey}, {Scarlata}, \&
  {Dowd}}]{Jaskot2017}
{Jaskot}, A.~E., {Oey}, M.~S., {Scarlata}, C., \& {Dowd}, T. 2017, \apjl, 851,
  L9, \dodoi{10.3847/2041-8213/aa9d83}

\bibitem[{{Kelson} {et~al.}(2000){Kelson}, {Illingworth}, {van Dokkum}, \&
  {Franx}}]{Kelson2000}
{Kelson}, D.~D., {Illingworth}, G.~D., {van Dokkum}, P.~G., \& {Franx}, M.
  2000, \apj, 531, 159, \dodoi{10.1086/308445}

\bibitem[{{Krumholz} \& {Matzner}(2009)}]{Krumholz2009}
{Krumholz}, M.~R., \& {Matzner}, C.~D. 2009, \apj, 703, 1352,
  \dodoi{10.1088/0004-637X/703/2/1352}

\bibitem[{{Krumholz} {et~al.}(2017){Krumholz}, {Thompson}, {Ostriker}, \&
  {Martin}}]{Krumholz2017}
{Krumholz}, M.~R., {Thompson}, T.~A., {Ostriker}, E.~C., \& {Martin}, C.~L.
  2017, \mnras, 471, 4061, \dodoi{10.1093/mnras/stx1882}

\bibitem[{{Kumari}(2018)}]{Kumari2018}
{Kumari}, N. 2018, PhD thesis, University of Cambridge

\bibitem[{{Lochhaas} \& {Thompson}(2017)}]{Lochhaas2017}
{Lochhaas}, C., \& {Thompson}, T.~A. 2017, \mnras, 470, 977,
  \dodoi{10.1093/mnras/stx1289}

\bibitem[{{Lopez} {et~al.}(2011){Lopez}, {Krumholz}, {Bolatto}, {Prochaska}, \&
  {Ramirez-Ruiz}}]{Lopez11a}
{Lopez}, L.~A., {Krumholz}, M.~R., {Bolatto}, A.~D., {Prochaska}, J.~X., \&
  {Ramirez-Ruiz}, E. 2011, \apj, 731, 91, \dodoi{10.1088/0004-637X/731/2/91}

\bibitem[{{McCray} \& {Snow}(1979)}]{McCraySnow1979}
{McCray}, R., \& {Snow}, T.~P., J. 1979, \araa, 17, 213,
  \dodoi{10.1146/annurev.aa.17.090179.001241}

\bibitem[{Micheva {et~al.}(2017)Micheva, Oey, Jaskot, \& James}]{Micheva2017}
Micheva, G., Oey, M.~S., Jaskot, A.~E., \& James, B.~L. 2017, \apj, 845, 165,
  \dodoi{10.3847/1538-4357/aa830b}

\bibitem[{{Oey} {et~al.}(2017){Oey}, {Herrera}, {Silich}, {Reiter}, {James},
  {Jaskot}, \& {Micheva}}]{Oey2017}
{Oey}, M.~S., {Herrera}, C.~N., {Silich}, S., {et~al.} 2017, \apjl, 849, L1,
  \dodoi{10.3847/2041-8213/aa9215}

\bibitem[{{Puls} {et~al.}(2008){Puls}, {Vink}, \& {Najarro}}]{Puls2008review}
{Puls}, J., {Vink}, J.~S., \& {Najarro}, F. 2008, \aapr, 16, 209,
  \dodoi{10.1007/s00159-008-0015-8}

\bibitem[{{Ramachandran} {et~al.}(2019){Ramachandran}, {Hamann}, {Oskinova},
  {Gallagher}, {Hainich}, {Shenar}, {Sander}, {Todt}, \&
  {Fulmer}}]{Ramachandran}
{Ramachandran}, V., {Hamann}, W.~R., {Oskinova}, L.~M., {et~al.} 2019, \aap,
  625, A104, \dodoi{10.1051/0004-6361/201935365}

\bibitem[{{Ravindranath} {et~al.}(2020){Ravindranath}, {Monroe}, {Jaskot},
  {Ferguson}, \& {Tumlinson}}]{Ravindranath}
{Ravindranath}, S., {Monroe}, T., {Jaskot}, A., {Ferguson}, H.~C., \&
  {Tumlinson}, J. 2020, \apj, 896, 170, \dodoi{10.3847/1538-4357/ab91a5}

\bibitem[{{Roy} {et~al.}(1992){Roy}, {Aube}, {McCall}, \& {Dufour}}]{roy1992}
{Roy}, J.-R., {Aube}, M., {McCall}, M.~L., \& {Dufour}, R.~J. 1992, \apj, 386,
  498, \dodoi{10.1086/171035}

\bibitem[{{Roy} {et~al.}(1991){Roy}, {Boulesteix}, {Joncas}, \&
  {Grundseth}}]{roy1991}
{Roy}, J.-R., {Boulesteix}, J., {Joncas}, G., \& {Grundseth}, B. 1991, \apj,
  367, 141, \dodoi{10.1086/169609}

\bibitem[{{Silich} \& {Tenorio-Tagle}(2017)}]{Silich2017}
{Silich}, S., \& {Tenorio-Tagle}, G. 2017, \mnras, 465, 1375,
  \dodoi{10.1093/mnras/stw2879}

\bibitem[{Silich {et~al.}(2020)Silich, Tenorio-Tagle, Martínez-González, \&
  Turner}]{Silich2020}
Silich, S., Tenorio-Tagle, G., Martínez-González, S., \& Turner, J. 2020,
  \mnras, 494, 97, \dodoi{10.1093/mnras/staa705}

\bibitem[{{Silich} {et~al.}(2004){Silich}, {Tenorio-Tagle}, \&
  {Rodr{\'\i}guez-Gonz{\'a}lez}}]{Silich2004}
{Silich}, S., {Tenorio-Tagle}, G., \& {Rodr{\'\i}guez-Gonz{\'a}lez}, A. 2004,
  \apj, 610, 226, \dodoi{10.1086/421702}

\bibitem[{{Smith} {et~al.}(2006){Smith}, {Westmoquette}, {Gallagher},
  {O'Connell}, {Rosario}, \& {de Grijs}}]{Smith2006}
{Smith}, L.~J., {Westmoquette}, M.~S., {Gallagher}, J.~S., {et~al.} 2006,
  \mnras, 370, 513, \dodoi{10.1111/j.1365-2966.2006.10507.x}

\bibitem[{{Sukhbold} {et~al.}(2016){Sukhbold}, {Ertl}, {Woosley}, {Brown}, \&
  {Janka}}]{Sukhbold}
{Sukhbold}, T., {Ertl}, T., {Woosley}, S.~E., {Brown}, J.~M., \& {Janka}, H.~T.
  2016, \apj, 821, 38, \dodoi{10.3847/0004-637X/821/1/38}

\bibitem[{{Tenorio-Tagle} {et~al.}(1997){Tenorio-Tagle},
  {Mu{\~n}oz-Tu{\~n}{\'o}n}, {P{\'e}rez}, \& {Melnick}}]{Tenorio-Tagle1997}
{Tenorio-Tagle}, G., {Mu{\~n}oz-Tu{\~n}{\'o}n}, C., {P{\'e}rez}, E., \&
  {Melnick}, J. 1997, \apjl, 490, L179, \dodoi{10.1086/311025}

\bibitem[{{Tenorio-Tagle} {et~al.}(2007){Tenorio-Tagle}, {W{\"u}nsch},
  {Silich}, \& {Palou{\v{s}}}}]{Tenorio-Tagle2007}
{Tenorio-Tagle}, G., {W{\"u}nsch}, R., {Silich}, S., \& {Palou{\v{s}}}, J.
  2007, \apj, 658, 1196, \dodoi{10.1086/511671}

\bibitem[{Thompson {et~al.}(2015)Thompson, Fabian, Quataert, \&
  Murray}]{Thompson}
Thompson, T.~A., Fabian, A.~C., Quataert, E., \& Murray, N. 2015, \mnras, 449,
  147, \dodoi{10.1093/mnras/stv246}

\bibitem[{Thuan {et~al.}(2014)Thuan, Bauer, \& Izotov}]{Thuan2014}
Thuan, T.~X., Bauer, F.~E., \& Izotov, Y.~I. 2014, \mnras, 441, 1841,
  \dodoi{10.1093/mnras/stu716}

\bibitem[{{Tolstoy} {et~al.}(1995){Tolstoy}, {Saha}, {Hoessel}, \&
  {McQuade}}]{Tolstoy1995}
{Tolstoy}, E., {Saha}, A., {Hoessel}, J.~G., \& {McQuade}, K. 1995, \aj, 110,
  1640, \dodoi{10.1086/117637}

\bibitem[{{Vink} {et~al.}(2001){Vink}, {de Koter}, \& {Lamers}}]{vink2001}
{Vink}, J.~S., {de Koter}, A., \& {Lamers}, H.~J.~G.~L.~M. 2001, \aap, 369,
  574, \dodoi{10.1051/0004-6361:20010127}

\bibitem[{{Westmoquette} {et~al.}(2014){Westmoquette}, {Bastian}, {Smith},
  {Seth}, {Gallagher}, {O'Connell}, {Ryon}, {Silich}, {Mayya},
  {Mu{\~n}oz-Tu{\~n}{\'o}n}, \& {Rosa Gonz{\'a}lez}}]{Westmoquette}
{Westmoquette}, M.~S., {Bastian}, N., {Smith}, L.~J., {et~al.} 2014, \apj, 789,
  94, \dodoi{10.1088/0004-637X/789/2/94}

\bibitem[{{Yeh} \& {Matzner}(2012)}]{Yeh2012}
{Yeh}, S.~C.~C., \& {Matzner}, C.~D. 2012, \apj, 757, 108,
  \dodoi{10.1088/0004-637X/757/2/108}

\end{thebibliography}
\bibliographystyle{aasjournal}

\end{document}